\documentclass[pra,reprint,superscriptaddress,longbibliography]{revtex4-1}

\usepackage{amsmath}    
\usepackage{graphicx}   
\usepackage[lofdepth,lotdepth, caption=false]{subfig}
\usepackage{verbatim}   
\usepackage{color}      
\usepackage{hyperref}   
\usepackage{dcolumn}
\usepackage{amssymb}
\usepackage{xcolor}  
\usepackage{tabularx}

\newcommand{\lcco}{La$_{2-x}$Ca$_{1+x}$Cu$_{2}$O$_{6}$}
\newcommand{\lccot}{La$_{1.9}$Ca$_{1.1}$Cu$_{2}$O$_{6}$}
\newcommand{\lccof}{La$_{1.85}$Ca$_{1.15}$Cu$_{2}$O$_{6}$}

\def\newr{\color{black}}

\def\lsco{La$_{2-x}$Sr$_x$CuO$_4$}
\def\lbco{La$_{2-x}$Ba$_x$CuO$_4$}

\def\ybco{YBa$_2$Cu$_3$O$_{6+x}$}

\begin{document}

\title{Evidence for magnetic-field-induced decoupling of superconducting bilayers in La$_{2-x}$Ca$_{1+x}$Cu$_{2}$O$_{6}$}

\author{Ruidan Zhong}
\thanks{Present address: Department of Chemistry, Princeton University,
Princeton, New Jersey 08544, USA}
\affiliation{Condensed Matter Physics and Materials Science Division, Brookhaven National Laboratory, Upton, New York 11973, USA}
\affiliation{Materials Science and Engineering Department, Stony Brook University, Stony Brook, New York 11794, USA}
\author{J.~A.~Schneeloch}
\thanks{Present address: Department of Physics,
University of Virginia, Charlottesville, Virginia 22904, USA}
\author{Hang Chi}
\author{Qiang Li}
\author{Genda~Gu}
\author{J.~M.~Tranquada}
\email{jtran@bnl.gov}
\affiliation{Condensed Matter Physics and Materials Science Division, Brookhaven National Laboratory, Upton, New York 11973, USA}

\date{\today} 

\begin{abstract}

We report a study of magnetic susceptibility and electrical resistivity as a function of temperature and magnetic field in superconducting crystals of \lcco\ with $x=0.10$ and 0.15 and  transition temperature $T_{c}^{\rm m} = 54$~K (determined from the susceptibility).  When an external magnetic field is applied perpendicular to the CuO$_2$ bilayers, the resistive superconducting transition measured with currents flowing perpendicular to the bilayers is substantially lower than that found with currents flowing parallel to the bilayers.  Intriguingly, this anisotropic behavior is quite similar to that observed for the magnetic irreversibility points with the field applied either perpendicular or parallel to the bilayers.  We discuss the results in the context of other studies that have found evidence for the decoupling of superconducting layers induced by a perpendicular magnetic field.

\end{abstract}

\maketitle
\section{introduction}

An unusual state of matter has been observed in the underdoped regime of at least one cuprate superconductor family, in which application of a $c$ axis magnetic field (perpendicular to the CuO$_2$ planes) destroys the phase coherence between the planes but appears to leave the superconducting response within the layers unaffected.   This effect was first detected by Schafgans {\it et al.} \cite{scha10} in $c$-axis optical reflectivity measurements of the Josephson plasma resonance in \lsco, {\newr with related behavior observed in a careful study of anisotropic susceptibility \cite{drac12}.}  It was confirmed in \lbco\ (LBCO) with $x=0.095$ through measurements of in-plane and $c$-axis resisitivity \cite{wen12}, with the decoupling still prominent in fields up to at least 35~T \cite{steg13}.  In these systems, the decoupling is correlated with the occurrence of charge-stripe order \cite{huck11,crof14,tham14} and presumably reflects a field-induced frustration of interlayer Josephson coupling by pair-density-wave superconductivity \cite{hime02,berg07,lee14,frad15} as proposed for the case of LBCO with $x=0.125$ \cite{li07,tran08}.

It is of interest to test whether such phenomena may occur in other cuprates.  A field-induced charge-density-wave transition has been observed in \ybco\ \cite{wu11,gerb15,chan16}.  It appears to be associated with the loss of superconducting order \cite{vign12}; however, torque magnetometry \cite{yu16} and specific heat \cite{rigg11} studies suggest that significant superconducting correlations survive to higher fields.  The relationship to the decoupling behavior described above remains to be resolved.

\begin{figure}[t]
\begin{center}
\includegraphics[width=0.45\textwidth]{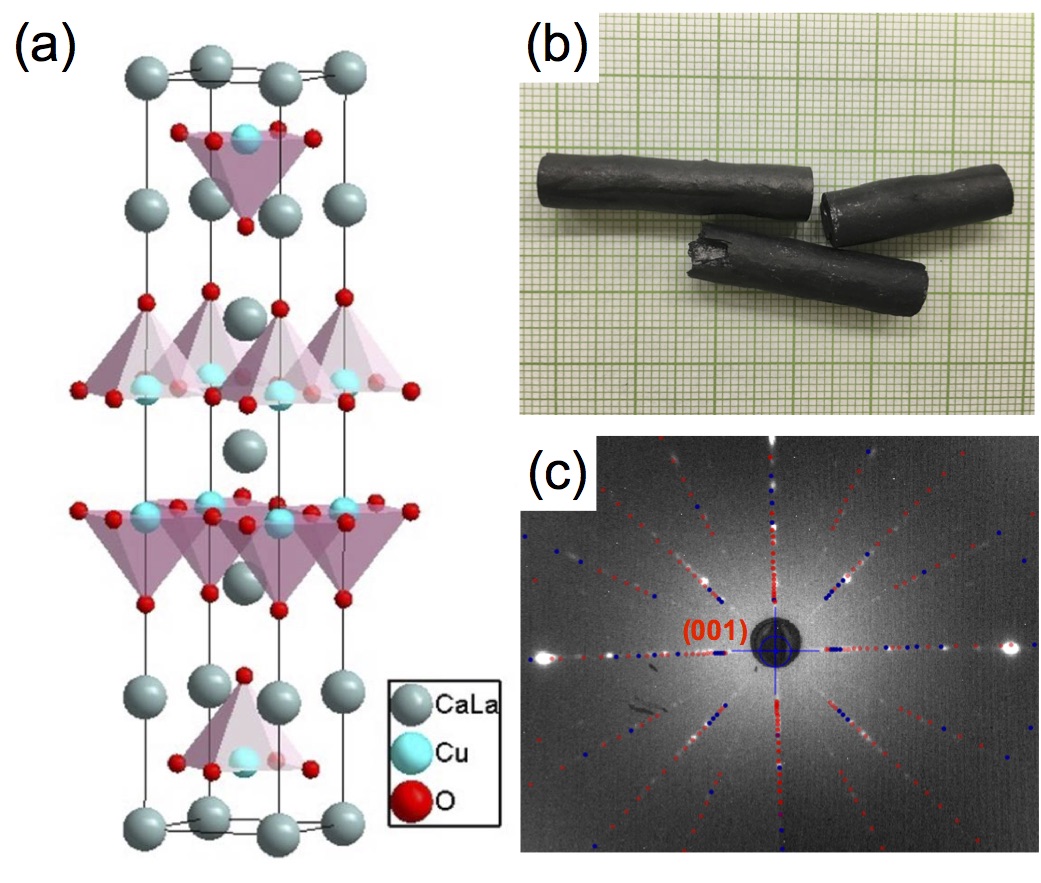}
\caption{(a) Crystal structure of the \lcco. (b) As-grown \lcco\ single-crystal rods grown by the traveling-solvent floating-zone method. (c) Laue back-diffraction image of \lcco\ single crystal with (001) direction ($c$-axis) pointing along the X-ray beam. Color points in (c) correspond to the calculated diffraction pattern, generated by software {\tt QLaue} \cite{QLaue}.
\label{fg:struc}}
\end{center}
\end{figure}

In this paper, we investigate another cuprate system.  We have recently succeeded in preparing superconducting crystals of \lcco\ (La-Ca-2126) of sufficient size for inelastic neutron scattering experiments \cite{schn17}.  Here we present a study of the anisotropic transport and magnetic susceptibility for two of these samples.  This system is different from \lsco\ in that it contains CuO$_2$ bilayers, as in \ybco, though the bilayers are stacked in a centered fashion, similar to Bi$_2$Sr$_2$CaCu$_2$O$_{8+\delta}$, as shown in Fig.~\ref{fg:struc}(a).   Interpretation of the results requires some care, as the high-pressure oxygen annealing essential to achieving superconductivity also results in the presence of minority phases of La$_{2-x}$Ca$_x$CuO$_4$ and La$_8$Cu$_8$O$_{20}$.  The latter compound is an antiferromagnetic insulator, while the former occurs as very thin intergrowths; all phases are crystallographically aligned along the $c$ axis \cite{schn17,hu14b}.  We believe that the segregation of these phases enhances the Ca concentration of the main La-Ca-2126 phase, thus providing sufficient hole doping to yield superconductivity.

Despite the complications, we find that application of a $c$-axis magnetic field results in distinct temperatures at which the in-plane and $c$-axis resistivities reach the normal-state.  The anisotropy in these transitions is quite similar to that in the irreversibility field for magnetic susceptibility measured with the field applied perpendicular or parallel to the $c$ axis.  We interpret the resistivity results as evidence of decoupling of the superconducting bilayers, indicating that such behavior is not unique to single-layer cuprates.

\section{Experimental Methods}

Crystals of La-Ca-2126 with $x=0.10$ and 0.15 were grown by the travelling-solvent floating zone method \cite{gu06a}.   The as-grown crystals were not superconducting, but they were converted to bulk superconductors by high-pressure annealing in a gas of 20\%\ O$_2$ / 80\%\ Ar \cite{schn17}.   Transmission electron microscopy (TEM) on an early version of such an annealed crystal demonstrated the presence of thin intergrowths of La$_{2-x}$Ca$_x$CuO$_4$ (La-214) \cite{hu14b}.  Layers of La-214 were observed with thicknesses of 1.5 or 3 unit cells along the $c$ axis, which are commensurate with 1 or 2 unit cells, respectively, of the La-Ca-2126 phase.  Neutron diffraction confirmed the presence of the thin La-214 layers, but, in combination with muon spin rotation, provided evidence for thicker layers of antiferromagnetic La$_8$Cu$_8$O$_{20}$ \cite{schn17}.   The volume fraction associated with the superconducting La-Ca-2126 was estimated to be $\sim70$\%.

The annealed single crystals were aligned to the desired orientations via X-ray Laue back-diffraction; the $ab$-plane diffraction pattern with the $c$-axis pointing along the incident X-ray beam is shown in Fig.~\ref{fg:struc}(c). Then the crystals were cut and polished into a nearly rectangular parallelepiped shape, with dimensions along $a\times b \times c$ of either {\newr (a)} $4\times1\times1$~mm$^3$ or {\newr (b)} $1\times1\times3$~mm$^3$.  For magnetization measurements, the field was applied along the long axis.  For resistivity measurements, the current was applied along the long axis, with the field always along the $c$ axis.  {\newr For each composition, at least two crystals were prepared of type (a), and they yielded similar results.  Because of difficulties preparing crystals without cracks, only one each of type (b) was studied.}

To study the magnetization and transport anisotropy, we performed measurements on samples with both orientations. The dc magnetic susceptibility measurements were performed using a Magnetic Property Measurement System from Quantum Design, 
with a superconducting quantum interference device (SQUID) magnetometer. Electrical resistivity was measured using the in-line four-point configuration, with an excitation current of 1 mA, in a Quantum Design Physical Properties Measurement System (PPMS).


\begin{figure*}[t]
\begin{center}
\includegraphics[width=0.8\textwidth]{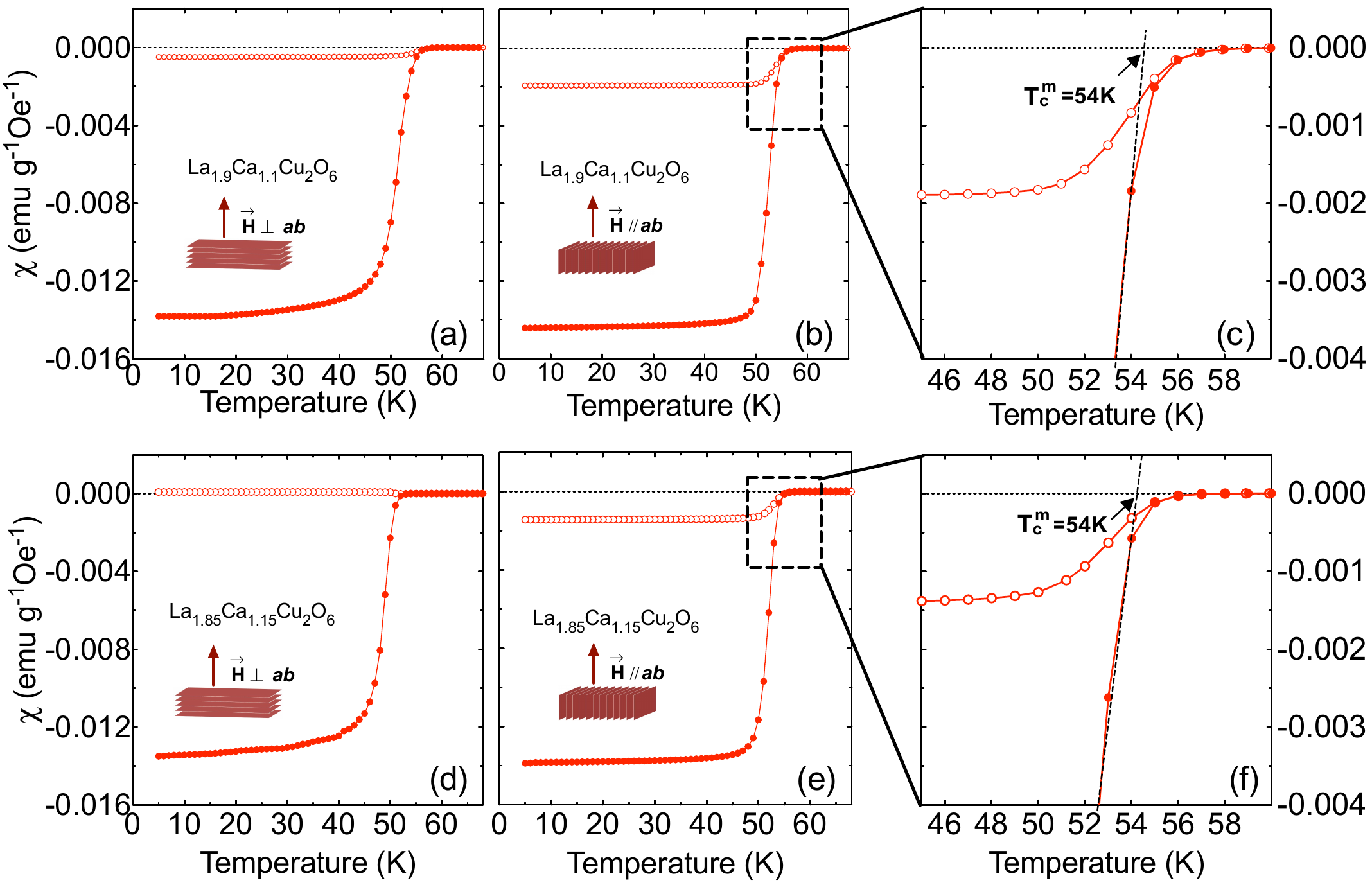} 
\caption{Mass magnetic susceptibility ($\chi=M/H$) of \lccot\ (top row) and \lccof\ (bottom row), measured with a magnetic field of 10~Oe applied (a), (d): perpendicular to the CuO$_{2}$ planes or (b-c), (e-f): parallel to the planes. Measurements were performed both in zero-field cooling (ZFC, filled circles) and field-cooling (FC, open circles) modes. The insets illustrate the relative directions of CuO$_{2}$ planes and the magnetic filed $H$. The criteria to determine $T_{c}^{\rm m}$ is shown in (c) and (f). 
\label{fg:lowf}}
\end{center}
\end{figure*}

\section{Susceptibility measurements}

Magnetic susceptibility measurements performed with a field of 10~Oe for both samples are displayed in Fig.~\ref{fg:lowf}.  Measurements with the field perpendicular [parallel] to the CuO$_2$ planes are shown in Fig.~\ref{fg:lowf}(a) and (d) [Fig.~\ref{fg:lowf}(b) and (e)].
Evidence for the {\newr compositional uniformity} 
of the superconducting phase in the annealed samples is given by the sharp superconducting transitions (width $\sim 5$ K).
A linear extrapolation of the transition region yields a magnetically-determined  superconducting transition temperature $T_{c}^{\rm m}= 54$~K for both compositions, slightly higher \footnote{The results are consistent with those previously presented for the same samples in Ref.~\cite{schn17} and characterized as $T_c=55$~K.  We have used a slightly different determination of $T_c$ here.} than previously reported single-crystal results \cite{ishi91,okuy94,hu14b,shi17}. 
For the zero-field-cooling (ZFC) curves, 100\%\ volume shielding would correspond to $\chi=0.0126$~emu~g$^{-1}$~Oe$^{-1}$; the observed response is of this magnitude despite the fact that the volume fraction of the La-Ca-2126 phase is just 70\%\ \cite{schn17}.  The Meissner fraction, determined by the field-cooled (FC) measurements, is considerably smaller; such behavior in cuprates is common and is typically attributed to flux pinning \cite{okuy94}.

For comparison, magnetization measurements on a La-Ca-2126 crystal with $x=0.1$ previously grown and annealed by our group had a similar $T_c^{\rm m}$ of 53.5~K, but also showed a small enhancement of the diamagnetic response below 13 K \cite{hu14b,shi17}.  That contribution might be due to La-214 layers, which are also present in the current samples; however, no low-temperature jumps in the diamagnetism are apparent in Fig.~\ref{fg:lowf}.


\begin{figure}[t]
\begin{center}
\includegraphics[width=0.35\textwidth]{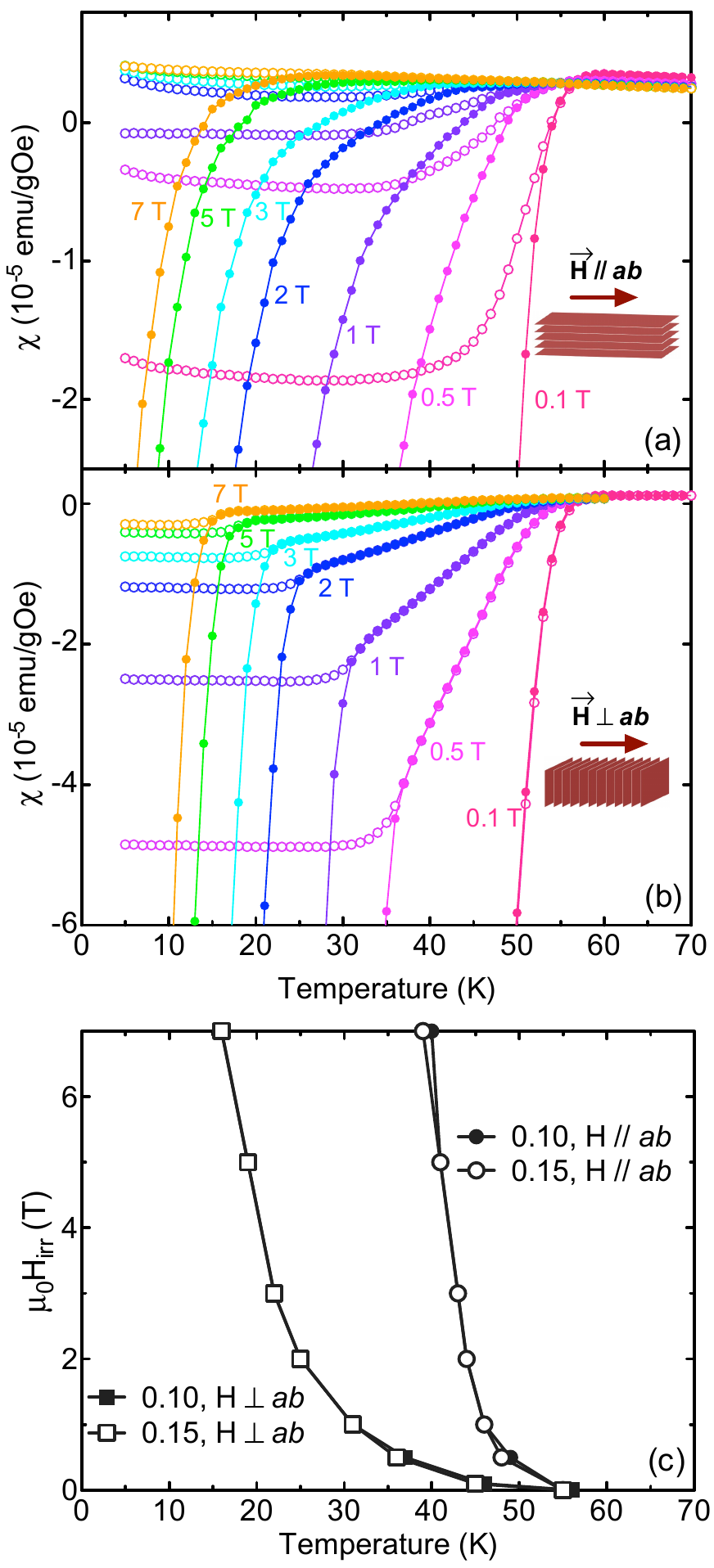} 
\caption{(a) The hysteretic curves of magnetic susceptibility versus temperature in sample \lccot\ under an in-plane field ($H\parallel ab$) up to 7 T. (b) Similar hysteretic curves obtained under a perpendicular field ($H\perp ab$). (c) The $T$ dependences of irreversibility field $H_{\rm irr}(T)$ in both directions relative to the CuO$_{2}$ planes for the \lcco\ ($x=0.10$ and 0.15) single crystals. 
\label{fg:highf}}
\end{center}
\end{figure}

We have also measured the temperature {\newr dependence} of $\chi=M/H$ for a range of magnetic fields up to 7~T.  A comparison of FC and ZFC results obtained for both field orientations on the $x=0.1$ sample is presented in Fig.~\ref{fg:highf}; the results for the $x=0.15$ sample (not shown) are quite similar.  
Hysteresis in the magnetic response provides evidence of pinning of magnetic vortices in the mixed phase \cite{sasa98,sasa00,li07b}. The irreversibility field, $H_{\rm irr}$, which is defined as the field that separates reversible and irreversible regions, provides a lower limit on the loss of static vortex matter.  It is observed to be sensitive both to temperature and to the orientation of the field with respect to the CuO$_2$ planes.  The temperature dependence of $H_{\rm irr}$ for both field orientations is summarized in Fig.~\ref{fg:highf}(c); note that the results for both samples are included and are nearly identical. As shown in the figure, the irreversibility field at any given temperature is much higher when the field is applied parallel to the planes.   In other words, it is easier to pin vortices aligned parallel with, and centered between, the superconducting bilayers than it is to pin vortices that pierce the bilayers {\newr \footnote{The layers of impurity phases may provide some local enhancement of pinning for $H\parallel ab$; however, it is not clear that this would significantly impact the irreversibility field for pinning within superconducting domains.  In any case, the anisotropy we find in $H_{\rm irr}$ is consistent with previous observations in cuprates free of impurity phases \protect\cite{sasa00,radz02,lund01}.}.}

It is important to avoid confusing $H_{\rm irr}$ with the loss of superconducting correlations.  As shown in a study of the reversible magnetization on a similar sample of La-Ca-2126 with $x=0.1$ \cite{shi17}, the temperature of the onset of diamagnetism shows very little change with fields up to 5~T, which is relatively close to what we observe for $H_{\rm irr}$ when the field is parallel to the bilayers.  To gain further insight into the behavior when the field is perpendicular, we turn to the measurements of transport anisotropy.


\section{Transport measurements}

\subsection{Zero-field results}

\begin{figure}[t]
\begin{center}
\includegraphics[width=0.4\textwidth]{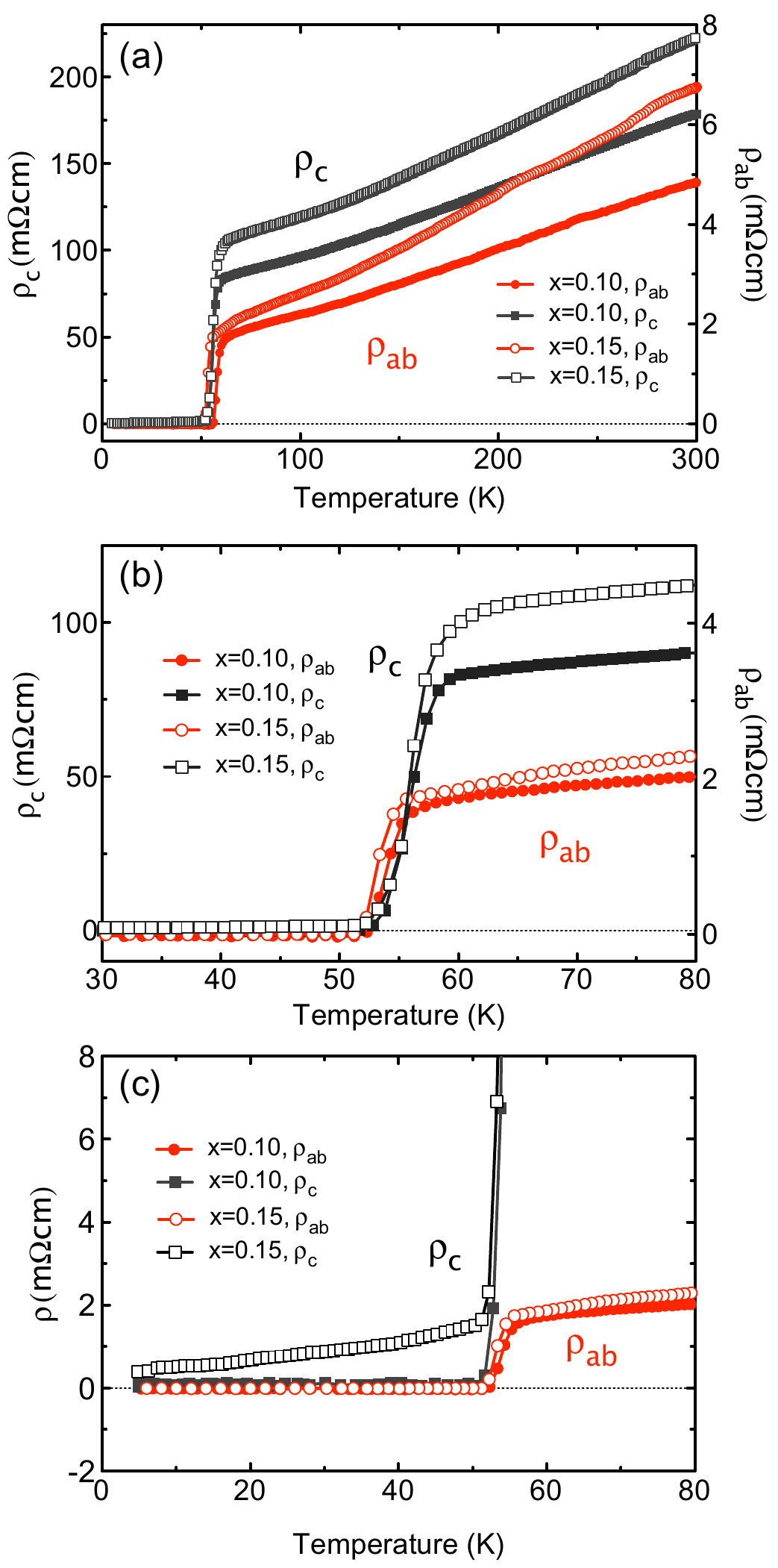} 
\caption{(a) The in-plane resistivity $\rho_{ab}$(T) and interlayer resistivity $\rho_{c}$(T) of the superconducting \lcco\ single crystals. Different $y$-axis values are used to illustrate the normal-state resistivity in both directions. (b) A zoomed-in figure showing the superconducting transition near 54~K. (c) A zoomed-in figure showing the finite resistivity in $\rho_c$ at low temperature. 
\label{fg:rhoz}}
\end{center}
\end{figure}

The resistivities measured with currents flowing in-plane, $\rho_{ab}$, and along the $c$-axis, $\rho_{c}$, are shown in Fig.~\ref{fg:rhoz}(a) for both samples.  We observed a metallic temperature dependence of the resistivity in both directions.  The magnitudes of $\rho_{ab}$ and $\rho_c$ are comparable to those reported for a superconducting crystal ($T_c\approx46$~K) of La-Ca-2126 with $x=0.1$ by Okuya {\it et al.} \cite{okuy94} (though they are each about an order of magnitude large than the results obtained on a flux-grown crystal with $x=0.13$ and $T_c\approx40$~K from an earlier study by Ishii {\it et al.} \cite{ishi91}).   Comparing with other cuprates, the magnitude of $\rho_c$ is comparable to that of optimally-doped \lsco\ \cite{naka93}, while lower than that of Bi$_2$Sr$_2$CaCu$_2$O$_{8+\delta}$ \cite{wata97} and higher than that of \ybco\ \cite{babi99}; however, $\rho_{ab}$ is about an order of magnitude larger than that for most cuprates \cite{naka93,wata97,ito93,babi99,ando04}.

The anomalously large magnitude of $\rho_{ab}$ requires further discussion.  We expect that the intrinsic $\rho_{ab}$ for our La-Ca-2126 samples should be similar to that of other cuprates near optimal doping.  The measured quantity, however, is impacted by the presence of other phases, especially the La$_8$Cu$_8$O$_{20}$ phase, which we expect to be insulating \cite{schn17}.  While the extra phases are coherently oriented with the main phase, the TEM study showed that interfaces between phases can occur along in-plane directions as well as along the $c$ axis.  As a consequence, the current path in a resistivity measurement may involve detours that mix contributions from in-plane and out-of-plane directions.  {\newr  Because $\rho_{ab}$ is much smaller than $\rho_c$, any current path that includes flow along the $c$ axis causes a substantial increase in the measured result for $\rho_{ab}$.}

These effects are also apparent when we look at an expanded view of the superconducting transition region, shown in Fig.~\ref{fg:rhoz}(b).  In particular, $\rho_{ab}$ has turned down by $T_c^{\rm m}$, but it only reaches zero at 52~K.  (In a single-phase sample, the situation is usually reversed, with the diamagnetism only starting to grow when the resistivity reaches zero.)

{\newr In contrast to $\rho_{ab}$, the magnitude of $\rho_c$ is found to be comparable to that of other cuprates.  We believe this is also compatible with an indirect current path.  If the current flowing along the $c$-axis of the La-Ca-2126 phase hits a layer of impurity phase, the path of least resistance may involve a detour parallel to the CuO$_2$ bilayers, before arriving at a domain where it can once again run along the $c$-axis of the main phase.  Any excursions within the planes are effectively like electrical shorts, and will not impact the measurement of $\rho_c$.

It is evident from Fig.~\ref{fg:rhoz}(c) that the measurements of $\rho_c$ cannot entirely avoid the insulating La$_8$Cu$_8$O$_{20}$ phase, as there is residual resistivity below $T_c$.  For the $x=0.15$ sample, $\rho_c$ just below $T_c$ is only $\sim1$\%\ of that in the normal state; considering that the resistivity of the insulating phase is much greater than that of La-Ca-2126 in the normal state, the relative path length through the insulator must be far less than 1\%.}
For the $x=0.10$ sample, $\rho_c$ drops to a much smaller but finite value at 52~K.

\subsection{Dependence on a $c$-axis field}

\begin{figure}[t]
\begin{center}
\includegraphics[width=0.4\textwidth]{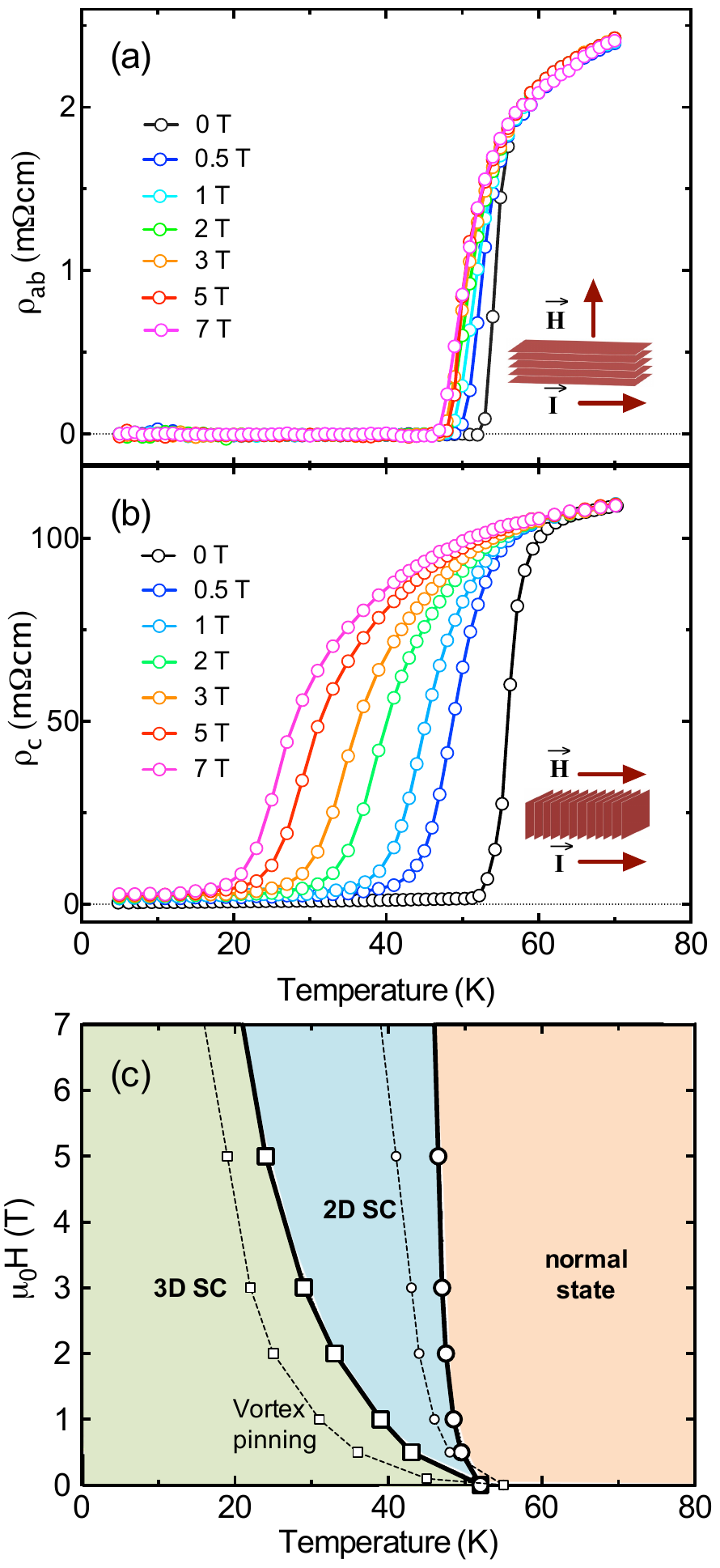} 
\caption{Electrical resistivity of (a) $\rho_{ab}$ and (b) $\rho_{c}$ as a function of temperature, obtained under a perpendicular field up to 7 T. (c) Phase diagram indicating the boundaries between the normal state, 2D and 3D superconductivity of \lccof.  {\newr Large squares (circles) represent the superconducting transition determined from $\rho_c$ ($\rho_{ab}$) with ${\bf H}\parallel {\bf c}$.  Small squares (circles) represent $H_{\rm irr}$ determined from $\chi$ with ${\bf H}\parallel {\bf c}$ (${\bf H}\perp {\bf c})$.}
\label{fg:rhof}}
\end{center}
\end{figure}

Next, we consider the impact of a $c$-axis magnetic field on the temperature dependence of $\rho_{ab}$ and $\rho_c$.  Measurements in fields up to 7~T are presented in Fig.~\ref{fg:rhof}.  As one can see, the field has only a small impact on the resistive transition as measured with currents parallel to the planes, whereas it causes a substantial reduction in the transition temperature when the measuring current is along the $c$ axis.

To quantify the anisotropic transition temperatures, we have to take account of the extrinsic resistivity observed in $\rho_c$ when $\rho_{ab}$ goes to zero.  To do this, we identify the transition in $\rho_c(T,H)$ as occurring when it reaches the value of $\rho_c(T=52~{\rm K},H=0)$.  With this choice, we obtain the effective transition temperatures plotted in Fig.~\ref{fg:rhof}(c).

We note that there is a substantial range of temperature and field in which the bilayers appear to be superconducting while there is no coherent superconducting transport between bilayers.  Analysis of the  magnetization in the reversible range near the onset of diamagnetism found evidence of 2D scaling behavior \cite{shi17}.  Our results indicate that this is not just a matter of strong anisotropy in a 3D system, but that superconductivity actually develops within bilayers before Josephson coupling can provide 3D coherence.

\section{Discussion}

The observed field-induced decoupling of superconducting layers is similar to the field-induced behavior found previously in \lbco\ with $x=0.095$ \cite{wen12,steg13}.  Of course, the latter was similar to the zero-field observation of decoupled superconducting planes in \lbco\ with $x=0.125$ \cite{li07,tran08}.  In LBCO, the superconductivity coexists with spin and charge stripe order \cite{huck11}, and it has been proposed that a suppression of Josephson coupling between layers is the consequence of pair-density-wave (PDW) superconducting order \cite{berg07,berg09b}.

{\newr  Could the presence of intergrowths of impurity phases qualitatively impact our results?  When the magnetic field is applied parallel to the planes, insulating layers would allow the field to locally penetrate the sample; however, unless there are special pinning effects near a surface, this should not impact the response of thick domains of the La-Ca-2126 phase.  In any case, it would only impact $H_{\rm irr}$ measured with ${\bf H}\perp {\bf c}$.  The decoupling effect observed in the resistivity  occurs for ${\bf H}\parallel {\bf c}$, where there is no obvious way for the intergrowths to impact the  anisotropic superconducting transition temperatures.}

Inelastic neutron scattering measurements have been performed on a large crystal of the La-Ca-2126 $x=0.15$ sample \cite{schn18}.  Although no static spin order was observed in association with the La-Ca-2126 phase, the bilayer magnetic excitations were found to be gapless, the latter being similar to the case of LBCO with $x=0.095$ \cite{xu14} and \lsco\ with $x=0.07$ \cite{jaco15}, and suggestive of intertwined order \cite{frad15}.  It would be interesting to test whether a $c$-axis magnetic field can induce spin order.

It is notable that the region of decoupled layers in Fig.~\ref{fg:rhof}(c) shows a significant correlation with the region between the irreversibility lines determined from the magnetic susceptibility measurements with the field parallel to $c$ or to the planes.  The irreversibility line in near-optimal Bi$_2$Sr$_2$CaCu$_2$O$_{8+\delta}$ for a $c$-axis field is also shifted far below the onset of diamagnetism, and there is evidence from in-plane and out-of-plane resistivity measurements for decoupling of the superconducting bilayers \cite{cho94,moro00}.  In that case, there is a substantial gap in the spin fluctuations below $T_c$ \cite{fong99,xu09}; however, charge-modulations are prevalent and enhanced by a magnetic field \cite{park10,dasi14,fuji14a,hoff02}.  The role of intertwined orders could provide a common connection \cite{frad15}.

\section{Summary}
In summary, we have presented experimental results of magnetic susceptibility and electrical resistivity measurements for superconducting single crystals La$_{2-x}$Ca$_{1+x}$Cu$_2$O$_6$ with $x=0.1$ and 0.15.  The magnetic susceptibility measurements confirm the bulk superconductivity in the samples, showing a narrow superconducting transition as well as $T_{c}^{\rm m}= 54$~K.  From the variation of the magnetization with magnetic field, we find that there is a large difference in the irreversibility line depending on whether the field is applied along the $c$ axis or in plane.
From the transport measurements, we find that the observed $\rho_{ab}$ is larger than expected, reflecting the impact of two non-superconducting minority phases that develop during the high-pressure annealing \cite{schn17}.  Nevertheless, we are able to identify the resistive superconducting transition and its dependence on the direction of current flow as a magnetic field is applied along the $c$ axis.  We find that there is relatively little shift in the transition temperature for currents flowing parallel to the planes, but a substantial change when the current flows between planes.  Hence, there appears to be a significant region in the phase space of temperature and magnetic field where we have superconducting bilayers that are decoupled from one another, and the boundaries are close to the anisotropic irreversibility lines.  These results demonstrate that the phenomenon of field-induced decoupling of superconducting layers is not limited to single-layer cuprates such as \lbco\ \cite{wen12,steg13} and they raise the question of the potential role of intertwined orders \cite{frad15}.

\section{acknowledgments}
Work at Brookhaven is supported by the Office of Basic Energy Sciences, Division of Materials Sciences and Engineering, U.S. Department of Energy under Contract No.\ {DE-SC}0012704. R.D.Z. and J.S. were supported by the Center for Emergent Superconductivity, an Energy Frontier Research Center. 

\bibliography{lno,theory,misc}

\end{document}